\documentclass[twocolumn,showpacs,preprintnumbers,amsmath,amssymb]{revtex4}
\usepackage{latexsym}
\usepackage{amssymb}
\usepackage{graphicx}% Include figure files
\usepackage{dcolumn}% Align table columns on decimal point
\usepackage{bm}% bold math
\newcommand{\W}{8cm}
\begin{document}

%\preprint{APS/123-QED}
\title{Influence of flow confinement on the drag force on a static cylinder}

\author{B. Semin}
\email{semin@fast.u-psud.fr}
\author{J.P. Hulin}%
\email{hulin@fast.u-psud.fr}
\author{H. Auradou}
\email{auradou@fast.u-psud.fr}
\affiliation{Univ. Pierre et Marie Curie-Paris 6, Univ. Paris-Sud, CNRS\\
Lab. FAST, B\^{a}t 502, Campus Univ., Orsay, F-91405, France.}

\date{\today}% It is always \today, today,

\begin{abstract}
The influence of confinement on the drag force $F$ on a static cylinder in a viscous flow inside
a rectangular slit of aperture $h_0$ has been investigated from experimental measurements
and numerical simulations.
At  low enough Reynolds numbers, $F$ varies linearly with the mean velocity
and the viscosity, allowing for the precise determination of  drag coefficients $\lambda_{||}$ and $\lambda_{\bot}$
corresponding respectively to a mean flow parallel and perpendicular to the cylinder length $L$.
In the parallel configuration, the variation of $\lambda_{||}$ with the normalized diameter $\beta = d/h_0$
of the cylinder is close to that for a $2D$ flow invariant in the direction of the cylinder axis and does not
diverge when $\beta = 1$. The variation of $\lambda_{||}$ with the distance
from the midplane of the model reflects the parabolic Poiseuille profile between the plates for
$\beta \ll 1$ while  it remains almost constant for $\beta \sim 1$.
In the perpendicular configuration, the value of $\lambda_{\bot}$ is close to that corresponding to
a $2D$ system only if $\beta \ll 1$ and/or if the clearance between the ends of the cylinder and the
side walls is very small: in that  latter case, $\lambda_{\bot}$ diverges as $\beta \rightarrow 1$ due
to the blockage of the flow. In other cases, the side flow between the ends of the cylinder and
the side walls plays an important part to reduce $\lambda_{\bot}$: a full $3D$ description of the flow
is needed to account for these effects.
\end{abstract}

%Valid PACS numbers may be entered using the \verb+\pacs{#1}+ command.

\maketitle
\section{Introduction}
Because of its fundamental and practical implications, considerable research efforts have
been devoted to the study of fluid flow past fixed or moving slender bodies.
Typical examples include the settling of suspensions of solid particles such as found in the paper industry
or the injection of fibers \cite{Wong2003,Yasuda2002}. Existing literature reports that, for flows in confined geometries such as pipes, hydrodynamic interactions between the fibers and the wall have a great influence on the fiber orientation and, in turn, on the flow properties \cite{Ku2008}. Before addressing such complex situations involving a large number of particles, simple models considering isolated fibers have recently been developed~\cite{benRichou2004}. In addition, the recent development of applications involving micro and nano rod-like objects~\cite{Czaplewski2004,Yesin2006} also raise questions on the hydrodynamic forces on objects placed in confined geometries such as microfluidic channels.

In this paper, the hydrodynamic forces $\bm{F}$ acting on a static cylindrical rod inside
a viscous flow in a slit of rectangular cross section $h_0 \times W$ (with $h_0\ll W$)
are determined both experimentally and numerically.
We have, in particular, compared the cases of a cylinder parallel and perpendicular to the flow.
The rod is a cylinder of high aspect ratio, \textit{i.e.} its length $L$ is always larger than
its diameter $d$ ($L \gg d$) but its length is of the same order as the slit size $W$. The effect of the confinement due to the two closest plane plates of the slit is particularly investigated: more precisely, the influence of the ratio $\beta=d/h_0$ is studied over a broad range of values up to $\beta \sim 1$ (very strong confinement). The influence of  the distance  $W$ between the two lateral sides  is also analyzed: it is particularly significant for cylinders normal to the mean flow and blocking it partly, resulting in large hydrodynamic forces.

For viscous flows (either confined or not),
the drag force $\bm{F}$ per unit length is proportional to
the dynamic viscosity $\eta$ and to the velocity $\bm{U}$ far from the
cylinder~\cite{GHP}. For unbounded flows, the leading term of the proportionality coefficients,
named in the following $\lambda _{||} = F_{||}/(\eta L U)$ (resp. $\lambda _{\bot} = F_\bot/(\eta L U)$) for flow
respectively parallel and perpendicular to the axis of the cylinder, is of
order $\epsilon=1/\ln(d/L)$ \cite{Batchelor1970,Cox1970}.

For slightly confined flows, wall correction terms increase the drag force.
The configuration in which a small cylinder sediments
half-way between parallel vertical plates separated by a large distance $h_0$ ($d \ll L \ll h_0 < W$) has been
studied experimentally by de Mestre~\cite{deMestre1973}. In this case, one has
$\lambda=\alpha_1(\epsilon+(\alpha_2 \frac{L}{h_0} + \alpha_3)
\epsilon^2)$, where the  parameters $\alpha _i$ are constants,
depending only on the orientation of the cylinder which is either vertical
or horizontal.
This configuration has also been studied in the limit
of cylinders of length large compared to the aperture
($h_0 \ll L \ll W$). In this case, the influence of the
boundaries on the drag is dominant and $\lambda$
 depends solely on the ratio $\beta=d/h_0$ (and not on the cylinder length $L$) and scales like $1/\ln(\beta)$
 for $\beta \ll 1$~\cite{Takaisi1955,Harper1967,Katz1975}.

These theoretical
predictions were confirmed and  extended to large values of $\beta$ by $2D$ numerical
simulations~\cite{benRichou2005} and experimental measurements~\cite{Stalnaker1979,White1945,Bouard1986,Bouard1997}
in the case of  cylinders moving with their axis normal to the flow.
The situation where the cylinders are fixed have been also considered
in the limit of low~\cite{Dvinsky1987,benRichou2004,Champmartin2007}
and high~\cite{Zovatto2001,Juarez2000,Sahin2004} Reynolds numbers and
for flows of complex fluids~\cite{Zisis2002,Bharti2007}.
Experimental results are however scarce in this geometry: Dhahir and co authors~\cite{Dhahir1989} measured forces
 on a cylinder of low aspect ratio ($d \sim L$) for different fluid rheologies
while Rehimi {\it et al}~\cite{Rehimi2008} use a geometry  similar to ours,
but do not perform force measurements.

In the present experiments, forces are measured on a static cylinder in a long slit of rectangular section
where the flow takes places.  Both $\beta=d/h_0$ and $L/W$ are varied as well as
the distance of the axis of the  cylinder from the center plane of the slit; both cases
of a cylinder parallel and perpendicular to the mean flow are studied.
In the perpendicular case
particular attention is given to the influence of the flow between the ends of the
 cylinder and the sides of the slit (if  $L < W$): flow in this region is highly
  three dimensional and  $3D$ simulations are needed to estimate it.

The other, parallel, configuration has not been studied up to now to our knowledge. However, previous
authors studied  theoretically in the viscous regime~\cite{Happel-Brenner} and experimentally
 in the inertial regime~\cite{Frei2000} the related problem of the forces on a cylinder located
inside another, coaxial, one. In the viscous regime, the velocity of a cylinder falling inside  another one has
 also been investigated \cite{Park1995}.
In this parallel case, we measure the drag forces induced by the flow on cylinders of
 different diameters ($0.04\leq \beta = d/h_0 \leq 0.83$) and for different locations in the aperture
 of the slit. In order to extend the range of physical parameters investigated, $2D$ numerical simulations
are performed; they allow, in addition to discriminate between  the contributions of the pressure
  and viscous shear forces to the global measured drag force.

The experimental setup is presented in section~\ref{Experimental
setup} and the numerical method in section~\ref{Numerical method}. The experimental and
numerical results are reported in sections~\ref{results lambapara} and~\ref{results lambdaperp}
respectively for cylinders parallel and perpendicular to the flow.
%%%%%%%%%%%%%%%%%%%%%%%%%%%%%%%%%
\section{Experimental setup and procedure}
\label{Experimental setup}
%%%%%%%%%%%%%%%%%%
\begin{figure}
\includegraphics[width=\W]{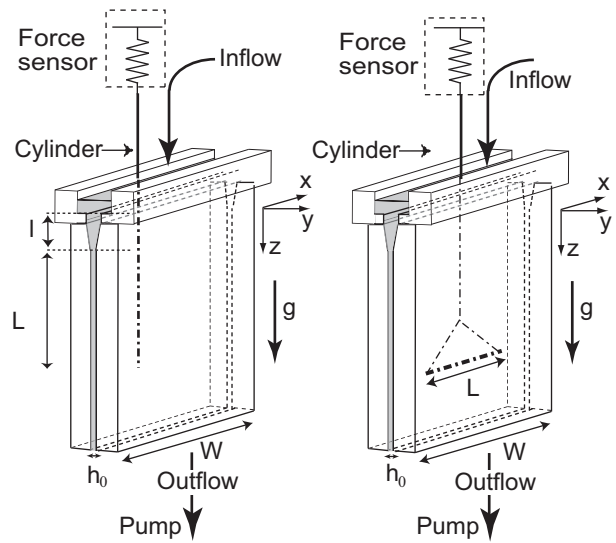}%
\caption{\label{fig:fig1}Left: schematic view of the experimental setup used to measure
$\lambda_{||}$, $L\in [49,200]$\,mm, $l=60$\,mm, $W=90$\,mm, $h_0$ is either
 $0.75$ or $4.9$\,mm.
 Right: setup for measuring $\lambda_\bot$, $h_0=4.9$\,mm, $L\in [20,89]$\,mm.}
\end{figure}
%%%%%%%%%%%%%%%%%%%
The experimental setups used for the determination of $\lambda_{||}$
and $\lambda_{\bot}$ are shown in Fig.~\ref{fig:fig1}:
they   consist of a slit of rectangular cross section placed
vertically and made of two transparent milled polymethyl methacrylate (PMMA)
plates.
The cell aperture has the constant value $h_0 = 4.9$ or $0.75\, \mathrm{mm}$
except in the upper $60$\,mm of the cell where it increases with height
from $0.75$\,mm (resp. $4.9$\,mm) to $5$\,mm (resp. $10$\,mm) for the first
(resp. second) model. This Y-shape profile was designed to ease
the insertion of the cylinders into the cell.

The cylinders
are hung on the hook of  computer controlled scales (Sartorius CP 225)
measuring drag forces of the flowing fluid on the cylinder in a
range of $10^{-7}$ to $8\times 10^{-1}\, \mathrm{N}$.
 A precision translation stage allows one to displace
the cylinders across the gap of the cell: in this way, the hydrodynamic
forces can be measured as a function of the distance from the walls.
Furthermore, the latter are transparent allowing one to determine precisely  the location
of the cylinders within the cell. For the cell with the largest aperture, side views
can also be obtained so that the distance separating the cylinder from the walls
may also be measured; this also allows one to control the parallelism of the object
with respect to the wall.

A gear pump sucks the fluid at the bottom side of the cell and
reinjects it into a bath covering the top side (the fluid flows therefore
always vertically downwards).
The fluids are either pure water or water-glycerol mixtures with a relative
 mass concentration of glycerol ranging from $50$ to $80\%$.

The value of the viscosity $\eta$  is determined by first measuring
 the density $\rho$ of the solutions
 and their temperature  before each series of experiments by means
of an Anton Paar $35$N densimeter. The viscosity $\eta$ is then determined
from the measured density and  temperature  by
interpolating tabulated values. The final uncertainty on the
value of the viscosity is about $3\%$ and is lower for all other parameters.
The relative influence of viscous and inertial effects is
characterized by the Reynolds number $Re=\rho h_0 U/\eta$ in which
 $U$ is the mean velocity far from the cylinder.

  %%%%%%%%%%%%%%%%%%%%%%%%%%%%%%%%%
\begin{table}
\caption{\label{tab:table1}Experimental parameters corresponding to the
measurements of $\lambda_{||}$. $d,\,L$: diameter and length of the cylinder;
$h_0$: cell aperture in the constant aperture region; $\beta=d/h_0$; $\eta$:
dynamic viscosity of the solutions. The symbols characterizing the different
experiments are the same as in the experimental figures.
For data corresponding to symbols ($\Box$), ($\boxplus$) and ($\boxtimes$),
$L$ is the difference between the lengths of two cylinders.}
 %%%%%%%%%%%%
\begin{ruledtabular}
\begin{tabular}{ccccccc}
 &$d$(mm)&$h$(mm)&$\beta$&$L$(mm)&$\eta$(mPa.s)&symbol\\
\hline
glass&1.5&4.9&0.31&89&40.0&$\Box$\\
PMMA&3.2&4.9&0.65&110&22.4&$\boxplus$\\
PMMA&4.05&4.9&0.83&100&21.7&$\boxtimes$\\
\hline
optical fiber&0.14&4.9&0.029&195&35.0&$\lhd$\\
polyester&0.20&4.9&0.041&177&17.5&$\Diamond$\\
polyester&0.20&4.9&0.041&177&7.34&$\triangledown$\\
silk&0.45&4.9&0.092&153&37.2&$\rhd$\\
glass&1.5&4.9&0.31&151&6.5&$\triangle$\\
glass&1.5&4.9&0.31&138&40.0&$\bigcirc$\\
iron&2.0&4.9&0.41&177&32.0&$\bowtie$\\
iron&4.0&4.9&0.82&184&37.6&$\odot$\\
PMMA&4.05&4.9&0.83&179&21.7&$\otimes$\\
polyester&0.18&0.75&0.24&135&119&$\blacktriangledown$\\
silk&0.45&0.75&0.6&82&24.0&$\blacktriangleright$\\
\end{tabular}
\end{ruledtabular}
\end{table}
%%%%%%%%%%%%%%%%%%%%%%%%%%%%%%%%%%
Measurement of $\lambda_{||}$ were performed for several cylinders
with different diameters (see Tab.~\ref{tab:table1}), which were
either rigid (glass, copper, iron, PMMA) or flexible (polyester or silk
threads).
The flexible threads were stretched prior to the experiments in order to remove
their residual curvature. These threads include multiple fibers so
that their diameter is not constant and varies periodically. Yet,
such variations were found to have a negligible effect on the drag
force: in the next parts, these threads are characterized
by their mean diameter.

%%%%%%%%%%%%%%%%%%%%%%%%%%%%%%%%%%%%%%%
\begin{table}
\caption{\label{tab:table2}Values of the experimental parameters
corresponding to the measurement of $\lambda_{\bot}$. All these
experiments were performed on the cell of aperture $h_0=4.9$\,mm and
width $W=90$\,mm; $Re<5$.}
%%%%%%%%%%%%%%%%%%%%%%%%%%%%%%%%%%%%%%%%%%%%%%%%%%
\begin{ruledtabular}
\begin{tabular}{cccccc}
 &$d$(mm)&$\beta$& range of $L$(mm)&symbol&position\\
\hline
 steel&$0.98$&$0.20$&$85$&$\blacktriangledown$ &center\\
 glass&$1.5$&$0.31$&$37.5 - 88$&$\blacktriangle$&center\\
 brass&$2.96$&$0.60$&$86.5$&$\blacksquare$&center\\
 PMMA&$3.2$&$0.65$&$44.7-88$&$\blacklozenge$&center\\
 PMMA&$4.05$&$0.83$&$67.5-88.7$&$\bullet$&center\\
 glass&$1.5$&$0.31$&$37.5 - 84$&$\triangle$&wall\\
 brass&$2.96$&$0.60$&$86.5$&$\square$&wall\\
 glass&$4.15$&$0.85$&$25.4 - 88.8$&$\bigcirc$&wall\\
 glass&$4.15$&$0.85$&$57.6$&$+$&wall\\
\end{tabular}
\end{ruledtabular}
\end{table}
%%%%%%%%%%%%%%%%%%%%%%%%%%%%%%%
For measuring $\lambda_{\bot}$ (see Tab.~\ref{tab:table2}),
the rigid cylinders are placed horizontally in the  cell of largest aperture
({\it{i.e.}} $h_0=4.9$\,mm); their center is halfway between the side walls of the cell.
There are hung using threads of small diameter ($100\,\mu\mathrm{m}$)
as shown in the right drawing of Fig.~\ref{fig:fig1}. Therefore, the drag forces  on the
threads and on the vertical rod add up: the former is however generally small compared to
the latter. Moreover, this extra force may be estimated numerically
(see Sec.~\ref{Numerical method}) and subtracted from the measurements.

The flow rate is increased stepwise from to $Q=0$ up to $Q=400\,\mathrm{mL/min}$ (resp.
$Q=1400\,\mathrm{mL/min}$) for the water glycerol mixtures (resp. for water), and
then decreased down to $Q=0$. Three such cycles are performed in order to
verify the reproducibility of the measurement. Figures
\ref{fig:fig3} and \ref{fig:fig4}
display the variations
of the drag force (averaged during each constant flow rate step)
as a function of the mean velocity $U$.
All data points corresponding to a same value of
$U$ almost coincide: this demonstrates
the very good reproducibility of the measurements and the lack of
hysteresis between the phases during which
the flow rate is increased or decreased.
%%%%%%%%%%%%%%%%%%%%%%%%%%%%%%%%%%%%
\section{Numerical simulation procedure}
\label{Numerical method}
%%%%%%%%%%%%%%%%%%%%%%%%%%%
\begin{figure}
\includegraphics[width=\W]{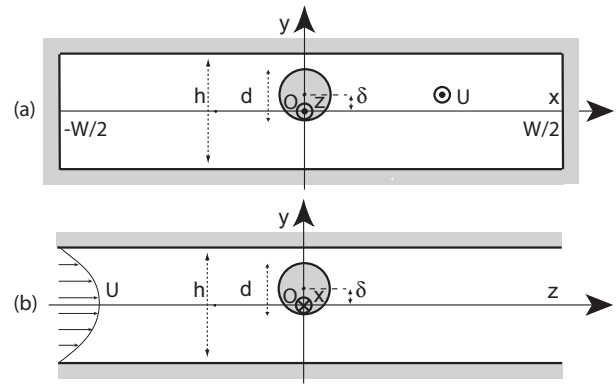}
\caption{\label{fig:fig2}Schematic $2D$ representation of the experimental configuration. (a): parallel case, (b): perpendicular case}
\end{figure}
%%%%%%%%%%%%%%%%%%%%%%%%%%%%%
In order to determine numerically $\lambda_{||}$ and
$\lambda_{\bot}$, the Stokes and incompressibility equations must be solved with appropriate
 boundary conditions. The shear stress and the pressure at the surface of the cylinder are then
  computed and added in order to determine the total hydrodynamic force.
 For flow parallel to the cylinders, $2D$ simulations provide reliable results;  in the perpendicular configuration,
 $2D$ simulations have also been used but $3D$ simulations are more realistic in several cases.
Figures \ref{fig:fig2} displays the geometry used in the $2D$ simulations:
the pressure gradient is applied in the $z$ direction, which is along the cylinder axis for $\lambda_{||}$ (see Fig.~\ref{fig:fig2}a))
and perpendicular of the cylinder axis for $\lambda_{\bot}$ (see Fig.~\ref{fig:fig2}b).
%%%%%%%%%%%%%%%%%%%%%%
 \subsection{Computation of $\lambda_{\parallel}$}
 \label{lambdaparallel}
%%%%%%%%%%%%%%%%%%%%%
For a constant pressure gradient parallel to the $z$ axis, the velocity
 is everywhere parallel to $z$ and $\bm{V}  =  V(x,y) \bm{e_z} $ due to the translational symmetry
 of the system.
 The governing equation of the flow then reduces
 to a $2D$ Laplace equation, which has been solved
by means of  the finite element program FreeFem++\cite{FreeFem++}. The
grids contain at least $12000$ nodes (values of the force computed with a finer mesh
size were identical to within $2\%$). The
hydrodynamic force per unit length $f$ is computed by adding to the shear force the
contribution of the pressure ($\frac{\pi}{4} \beta^2 \partial p /\partial z$ , where $\beta=d/h_0$).
The influence of the lateral side plates is taken into account by setting a zero-velocity
 boundary condition for $x = \pm W/2$.
A cell of infinite width is modeled by assuming, for $x = \pm W/2$, a parabolic variation of
the velocity with the distance $y$ from the midplane.

The force measured experimentally is the integral of the forces acting along
the full length of the cylinder. Assuming that the velocity profile
in the gap of the cell depends only on its local aperture $h(y)$ (lubrication approximation),
the total force is computed numerically from:
%%%%%%%%%%%%%%%
\begin{equation}
F_{lub.}(L)=\int_{-l}^{L} f(z) \mathrm{d}z, \label{eq:Flub}
\end{equation}
%%%%%%%%%%%%%%%%%%
in which $L$ is the length of the part of the cylinder located inside the constant aperture
domain (See Fig.~\ref{fig:fig1}) and $l$ corresponds to the part inside the Y-shaped section
(in the upper fluid bath, the fluid velocity is small enough so that its contribution to
the force can be neglected).
%%%%%%%%%%%%%%%%%%%%%%%%%
\subsection{Computation of $\lambda_{\bot}$}
\label{lambdaperpendicular}
%%%%%%%%%%%%%%%%%%%%%%%%
\subsubsection{$2D$ computation}
%%%%%%%%%%%%%%%%%%%%%%%%%%
Here, we consider the configuration in which the flow is normal to
the cylinder (see Fig.~\ref{fig:fig2}b).
 Far from the cylinder, the velocity has a parabolic Poiseuille profile.
In this $2D$ approach, the flow is assumed to be invariant in the $x$ direction so that:
 $\bm{V}  =  V_y(y,z) \bm{e_y} +V_z(y,z) \bm{e_z}$.
The $2D$ Stokes equation is solved numerically by means of the Freefem++ program,
using a mesh containing at least $15000$ nodes. The length
of the computational domain in the $z$ direction is $8$ times its
height in the $y$ direction (we  checked that choosing a longer computational domain in the $z$
 direction has a negligible influence on the value of the force). The shear and the pressure
forces are then computed from the velocity field.

In order to validate the present method, we compared its predictions in the particular case
of a cylinder located halfway between the plates to those available in the literature. In particular,
Richou et al.~\cite{benRichou2004} estimated
analytically an asymptotical value of $\lambda_{\bot}$  in the limit when the free
space between the cylinder and the walls is small ($\beta \approx 1$) by using the  lubrication approximation.
This approximation is valid in regions where one of the front walls  and the surface of the cylinder are close to
each other and nearly parallel. Since the volume flow rate is conserved along the stream-tubes,
these are the same regions in which the fluid velocity is the highest (and therefore also the friction force and
the pressure gradient). Rewriting these results~\cite{benRichou2004} with our notations gives:
%%%%%%%%%%%%%%%%%%%%%
\begin{eqnarray}
\lambda_{\bot}(\beta,\delta=0) \underset{\beta \rightarrow 1}{\longrightarrow} \nonumber \\
 \underbrace{6\pi \sqrt{2}\frac{{\beta}^{1/2}}{{(1-\beta)}^{3/2}}}_{\lambda_{\mathrm{shear}}}+
 \underbrace{9\pi \sqrt{2}\frac{{\beta}^{3/2}}{{(1-\beta)}^{5/2}}}_{\lambda_{\mathrm{pressure}}},
\label{analytical value of lambdaperp}
\end{eqnarray}
%%%%%%%%%%%%%%%%%%%%%%
where $\delta$ is the distance between the middle of the cell and the axis of the cylinder (See Fig.~\ref{fig:fig2}b).
The pressure term is dominant when $\beta  \longrightarrow 1$.
Still using the lubrication approximation, we extended this result to the case of a cylinder touching
the wall in the constant aperture region (thickness of the free space equal to zero on one of the sides) with:
%%%%%%%%%%%%%%%%
\begin{eqnarray}
\lambda_{\bot}(\beta,\delta=(h_0-d)/2)  \underset{\beta \rightarrow
1}{\longrightarrow} \frac{1}{2\sqrt{2}}
\lambda_{\bot}(\beta,\delta=0).\label{analytical value of lambdaperp wall}
\end{eqnarray}
%%%%%%%%%%%%%%%%%%
Table~\ref{tab:table3} compares, for different values of $\beta = d/h_0$, the results
of the present $2D$ numerical simulations
to the predictions of Eq.~(\ref{analytical value of lambdaperp}) and to the theoretical
and numerical results of other authors for a cylinder located
halfway between the plates. The good agreement between the different values
validates the present numerical procedure.
%%%%%%%%%%%%%%%%%%%%%%%%%%%%
\begin{table}
\caption{Comparison between values of $\lambda_\bot$ for a cylinder halfway between
the plates from the present work and from previous numerical and theoretical studies.
The analytical solution of ref.~\cite{benRichou2004} uses
Eq.~(\ref{analytical value of lambdaperp});  all the results of
the table correspond to  $2D$ systems.}
%%%%%%%%%%%%%%%%%%%%%%%%%%%%%%%%%%
\label{tab:table3}
\begin{ruledtabular}
\begin{tabular}{cccccc}
& Numerical&  Numerical& Analytical & Analytical \\
 $\beta$&Present work&Richou(2004)&Faxen (1946)&Richou(2004)\\
\hline
0.01&   5.100&  5.309&  5.109&\\
0.1&    13.34&  13.74&  13.36&  11.52\\
0.4&    72.69&  73.28&  72.93&  72.55\\
0.6&    262.4&  266.8&  &   265.3\\
0.8&    1850&   1884&   &   1866\\
0.96&   $1.205\times 10^5$&    $1.149\times 10^5$&    &   $1.208\times 10^5$\\
0.99&   $3.955\times 10^6$&    $3.174\times 10^6$&    &
$3.965\times 10^6$\\
\end{tabular}
\end{ruledtabular}
\label{tab:table}
\end{table}
%%%%%%%%%%%%%%%%%%%%%%%%%%%%%%%%%%%%%
%%%%%%%%%%%%%%%%%%%%%%%%%%%%%%
\subsubsection{$3D$ computation}
%%%%%%%%%%%%%%%%%%%%%%%%%
The $2D$ approaches described above assume that the
cylinder is infinitely long: they describe correctely therefore only the case of a cylinder of
length equal to the width of the slit ($L=W$ in Fig.~\ref{fig:fig1}).
If $L < W$, there is a deviation of the flow lines from the $z = cst$ planes
(Fig.~\ref{fig:fig2}) resulting from the free space separating the edge of the cylinder
and the lateral walls. If this lateral clearance is large, this reduces strongly the
 force on the cylinder compared to the $2D$ configuration.

In order to estimate the magnitude of this latter effect and compare the results to the
experimental observations, we solved numerically the $3D$ Stokes equation in some of
our experimental configurations  by means of the Freefem3D program~\cite{FreeFEM3D}.
The numerical discretization consists in a P1-P1 finite element method using
penalty for the pressure stabilization; the number of vertices is at least $70000$.
The ratio between the numerical lengths in the directions $x$ and $y$ is $18$
(like in the experimental cell), and the ratio between the numerical lengths in the
 directions $y$ and $z$ directions is $12$.
%%%%%%%%%%%%%%%%%%%%%%%%%%%%%%%%%%%%%%%%%%%%%%%%%%%%%%%%%%%%
\section{Experimental and numerical variations of $\lambda_{||}$}
\label{results lambapara}
%%%%%%%%%%%%%%%%%%%%%%%%%%%%%%%%%%%%%%%%%%%%%%%%%%%%%%%
Most previous experimental studies investigated only configurations in which
the cylinders are weakly confined ({\it{i.e.}} $\beta \ll 1 $)
and/or located halfway between the walls of the channel. The present
experimental setup has allowed us to investigate the cases in which the
confinement is strong as well as the variations of the force when the cylinder
 comes close to the wall.
%%%%%%%%%%%%%%%%%%%%%%%%%%%%%%%%%%%%%%%%%%%%%%%%%%%%%
\subsection{Variation of the drag force on a cylinder parallel to the flow
with the mean velocity}
\label{force velocity}
%%%%%%%%%%%%%%%%%%%%%%%%%%%%%%%%%%%%%%%%%%%%%%%%%%%%%%%%%%%%%%%%%%%
In the present study, the dependence of the drag force on the geometry of the system for viscous flow is addressed. For this flow condition, the drag increases linarly with the mean flow velocity, so that $\lambda_{||}$ is independant of the velocity.
Therefore, before analyzing the dependence of $\lambda_{||}$ on the geometrical
parameters of the flow, we investigated the domain
 in which the force and
the velocity are proportional for fluids of different viscosities.

%%%%%%%%%%%%%%%%%%%%%%%%%%%%%%%%%%%%%%%%%%%%%%%%%%%%%%%
\begin{figure}[htbp]
\includegraphics[width=\W]{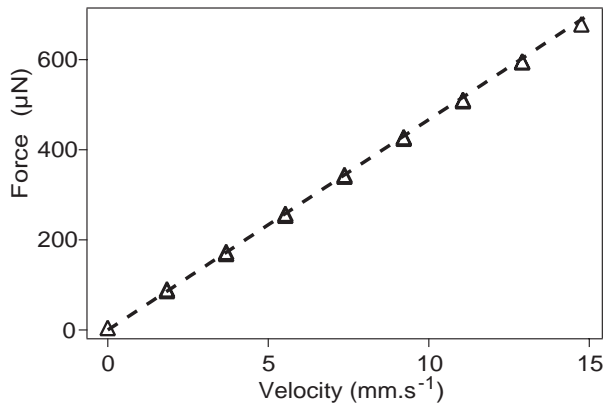}
\caption{\label{fig:fig3}Variation of the force on a vertical
cylinder located halfway between the plates as a function of the
mean velocity $U$ of a water glycerol mixture. $d=1.5$\,mm, $h_0=4.9$\,mm,  $\beta=0.3$
and $\eta=40.0$\,mPa.s,  Reynolds number $Re<2.2$.($\triangle$): experiments.
 Dashed line: numerical computation.}
\end{figure}
%%%%%%%%%%%%%%%%%%%%%%%%%%%%%%%%%%%%%%%%%%%%%%%%%%
Figure~\ref{fig:fig3} displays a set of measurements obtained
using a water-glycerol mixture as the flowing fluid: the dashed line is the variation
of the force with the velocity predicted by the $2D$ numerical simulations
of section~\ref{lambdaparallel}. The difference between the
experimental data and the numerical results is lower than $5\%$ for all
the experiments performed in the parallel case, without any adjustable parameter. The
linear increase of the force with the fluid velocity implies that
the viscous effects are dominant. This is in agreement with the low value of the Reynolds
number $Re$ which is less than $3$ for all the experiments  using the
water-glycerol solution including that of Fig.~\ref{fig:fig3}.

%%%%%%%%%%%%%%%%%%%%%%%%%%%%%%%
\begin{figure}[htbp]
\includegraphics[width=\W]{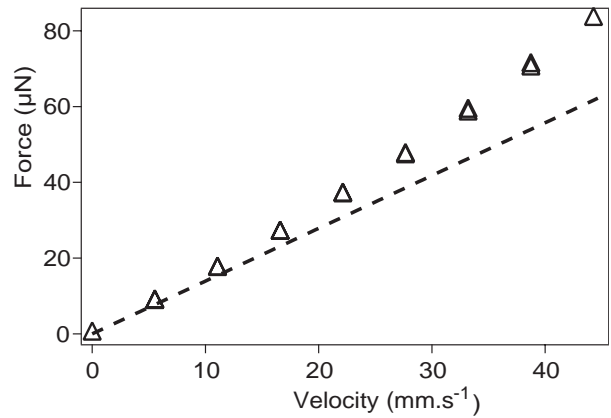}
\caption{\label{fig:fig4}Same measurements as in Fig. \ref{fig:fig3}
but using water ($\eta =1.0\,\mathrm{mPa.s}$): $Re \simeq 220$ for the highest
 flow rate. ($\triangle$): experiments. Dashed line: numerical computation.}
\end{figure}
%%%%%%%%%%%%%%%%%%%%%%%%%%%%%%%%%%%%%%%%%%%%%%%%%%%%%%%%%%%%%%%
This condition is not satisfied any more  for water: in this case, the
Reynolds number can be as high as $200$ and inertial effects are
important even at moderate velocities as can be seen in Fig.~\ref{fig:fig4}.
The force  follows at first a linear trend
 with $U$: moreover, the slope at the origin is in good agreement with the
 theoretical value (dashed line).
 However, and in contrast with the case of the water-glycerol solution,
the curve deviates from the linear trend at higher velocities which reflects the
 increase of the inertial effects: the deviation becomes large for mean velocities
 $U \ge 10\, \mathrm{mm.s}^{-1}$ corresponding to Reynolds numbers $Re \ge 50$.
In the following, only results corresponding to the
domain of linear variation of $F$ with $U$ are reported.
%%%%%%%%%%%%%%%%%%%%%%%%%%%%%%%%%%%%%%%%%%%%%%%%%%%%%%
  \subsection{Variation of $\lambda_{||}$ with the diameter}
  \label{lambdapara diameter}
%%%%%%%%%%%%%%%%%%%%%%%%%%%%%%%%%%%%%%%%%%%%%%%%%%%%%%
\begin{figure}
\includegraphics[width=\W]{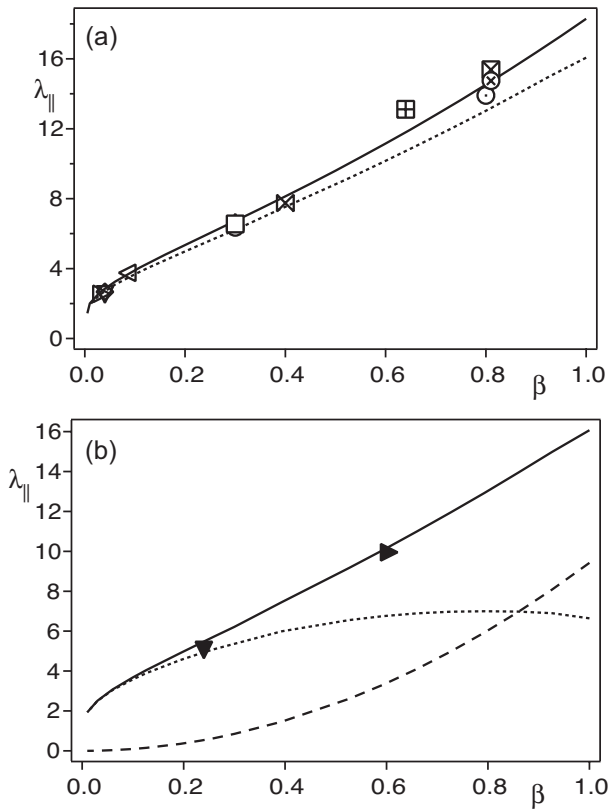}
\caption{\label{fig:fig6}Variation of $\lambda_{||}$ vs. $\beta = d/h_0$ for
cylinders located halfway between the walls. Symbols of experimental data
points are defined in Tab.~\ref{tab:table1}.
Solid lines: $2D$ numerical predictions. (a):  cell of aspect ratio $W/h_0 \approx 18$
($h_0=4.9\,\mathrm{mm}$). Dashed line: numerical value in the limit $W/h_0 \rightarrow \infty$.
(b): cells of very large aspect ratio $W/h_0\ge 120$ ($h=0.75\,\mathrm{mm}$);
dotted line: viscous shear force term, dashed line: pressure term.}
\end{figure}
%%%%%%%%%%%%%%%%%%%%%%%%%%%%%%%%%%%%%%%%%%%%%%%%%%%%%%
In this section, we study how confinement influences the hydraulic forces on cylinders
with their axis vertical ({\it i.e.} parallel to the flow) and located halfway between
the  two parallel vertical walls of spacing $h_0$. For this purpose, we determine
$\lambda_{||}$ from force measurements and we study its variation as a function of
the ratio $\beta = d/h_0$ by using cylinders of different diameters $d$.

However, in this configuration, the local distance between the front walls varies
continuously in the Y-shaped section at the top of the cell. As a result,
the drag force per unit length increases continuously along the rod and is only
 constant in the region of constant aperture $h_0$. In order to determine  the value of the drag specifically  in this latter region, we repeat the experiment  twice with  cylinders of two
different lengths $L_1$ and $L_2$ (large enough to reach the constant aperture
domain), but otherwise identical. From Eq.~(\ref{eq:Flub}), the difference between
 the forces measured
at a  same mean flow velocity $U$ on the two cylinders is:
$\Delta F = \lambda_{||}\, \eta\, U\,(L_1-L_2)$.

Figure~\ref{fig:fig6}a displays (solid lines) the  variation of $\lambda_{||}$
with $\beta$ predicted from the $2D$ numerical simulations,
for a cell of aspect ratio $W/h_0=18$ ({\it i.e.} of  aperture $h_0=4.9\,\mathrm{mm}$).
Values obtained by subtracting the  forces measured for two cylinders of different lengths
 (($\Box, \boxplus, \boxtimes$) symbols) are in good agreement with the numerical model.

In the present experiments, the part $l$  of the length  of the cylinder inside
the Y-shaped section is generally much
smaller than the length $L$ inside the constant aperture zone. In this case,
we estimate $\lambda_{||}$ from a single   measurement of the force $F$
by assuming that, although the local aperture $h(z)$ varies,
$\lambda_{||}$ is the same in all sections
and that the corresponding local mean velocity  is $U(z) = h_0\,U/h(z)$
(in order to insure mass conservation).
Eq.(\ref{eq:Flub}) becomes then: $F = \lambda _ {||}
\eta U (L+\int_{-l}^{0} (h/h(z)) \mathrm{d} z)$ with $h(z) =h_0 + (h_0-h_i)z/l $ ($h_i$ is the aperture at
the top of the Y-shaped section).
$\lambda_{||}$ is then related to the  force $F$ by: $\lambda_{||} = F/(\eta\,U\, L^*)$
in which  $L^*=L+ l\, h_0\, ln(\frac{h_i}{h_0})/(h_i-h_0)$ is an equivalent length.
All data points, except those corresponding to the ($\Box, \boxplus, \boxtimes$) symbols
in Figure~\ref{fig:fig6}a, were obtained by this ``equivalent length'' method:
its validity is demonstrated by the
small difference between  values obtained  for same experimental parameters
by the two  methods ($\otimes$
and $\boxtimes$ symbols).

Points ($\Diamond$) and ($\triangledown$) correspond to the same experimental configuration
 except for the viscosity which differs by a factor of $2$: they coincide
within experimental error which confirms  that the force is proportional
 to the viscosity.

In figure~\ref{fig:fig6}a, experimental data points are (as expected) closer to the numerical values taking into account
the finite aspect  ratio $W/h_0=18$ of  the cell (solid line) than to those assuming an infinite width $W$ (dashed line).
The difference between the two curves is however only of the order of $10\%$,
indicating that the effect of the lateral boundaries is weak.
This correction becomes completely negligible for the narrower cell ($h_0=0.75$ mm and $W/h_0=120$)  and the
results are then  the same as for $W/h_0 \rightarrow \infty$. In this case, it is more difficult to
control  experimentally the position of the cylinders: the two experimental values  obtained
(using the equivalent length approach)  are, however, in good
agreement with the numerical predictions (see Figure~\ref{fig:fig6}b).

Still in the high aspect ratio limit, we computed numerically
the pressure force term (dashed line in Fig.~\ref{fig:fig6}b)  and the shear force term
 (dotted line in Fig.~\ref{fig:fig6}b). On the one hand, for $\beta > 0.2$, the
viscous contribution levels off and decreases for $\beta > 0.8$; on the other
hand, the pressure contribution increases sharply when
$\beta$ increases. For $\beta > 0.2$, the sum of the two contributions ({\it i.e.}  $\lambda_{||}$)
increases almost linearly with $\beta$ ($\lambda_{||}\approx 2.1 + 13.8\, \beta$),
and does not diverge for $\beta=1$, due to the weak perturbation of the flow.
%%%%%%%%%%%%%%%%%%%%%%%%%%%%%
\subsection{Variation of $\lambda_{||}$ with the location of the cylinder in the aperture}
\label{lambdapara gap}
%%%%%%%%%%%%%%%%%%%%%%%%%%
%%%%%%%%%%%%%%%%%%%%%%%%%%%%%
\begin{figure}[htbp]
\includegraphics[width=\W]{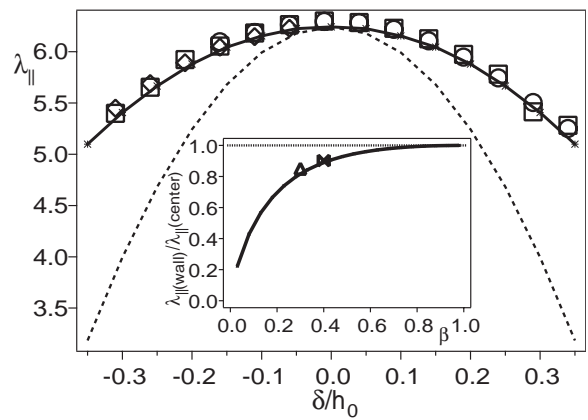}
\caption{\label{fig:fig7}Variation of $\lambda_{||}$ with the normalized distance
$\delta/h_0$ from the middle plane ($|\delta/h|<|0.5-d/(2h_0)|=0.35$) for
$d=1.5$\,mm and $h_0=4.9$\,mm. Other parameters are the same as for the ($\triangle$)
symbols in Tab.~\ref{tab:table1}. ($\Box$, $\Diamond$, $\circ$):  repeat experimental runs
for the same parameter values as ($\triangle$). Solid line: numerical computation of the force;
dashed line: Poiseuille  profile.
Insert: variation of the  ratio of the values of $\lambda_{||}$ at the wall and at the center
as a function of $\beta$. ($\triangle$) and ($\bowtie$): experimental values; solid line:
numerical computations.}
\end{figure}
%%%%%%%%%%%%%%%%%%%%%
The experimental setup also allows one to move precisely the cylinder
across the gap of the cell. The normalized  distance $\delta/h_0$ along $y$ of the axis of the cylinder
from the middle plane of the cell is then
determined with a precision better than $50\,\mu \mathrm{m}$ using the
side view pictures.
The experimental values of $\lambda_{||}$ are plotted as a function of $\delta/h_0$ in
Fig.~\ref{fig:fig7} : in this case, the sequence of displacements towards and away
from the walls is performed three times. For all the values of $\beta$
considered in the present work, this curve has a parabolic shape
with a maximum for $\delta/h_0 = 0$.
The coefficient $\lambda_{||}$ decreases when approaching the wall
 (solid line Fig.~\ref{fig:fig7}) but less than if it followed a parabolic variation for
 a Poiseuille velocity profile with no cylinder present (dashed line). The influence of the finite
diameter of the cylinder, particularly when it is of the order of the distance to the wall, accounts
for this difference.

The insert of Figure~\ref{fig:fig7}
displays the variation with $\beta$ of the ratio of the values of
$\lambda_{||}$ for a cylinder in contact with the walls and
in the middle of the cell. This curve is well fitted by the function
$1- 0.9\times \exp(-3.4 \beta)$ in the range $0.03\leq \beta \leq 1$. Even
though the overall variation is always parabolic, the decreasing trend of
$\lambda_{||}$ near the wall is stronger for lower values of $\beta$;
for $\beta \ll 1$, this variation  is expected to be similar to the Poiseuille
parabolic profile.
%%%%%%%%%%%%%%%%%%%%%%%%%%%%%%%%%%%%
\section{Experimental and numerical variations of $\lambda_\bot$}
\label{results lambdaperp}
%%%%%%%%%%%%%%%%%%%%%%%%%%%%%%%%%%%%
While the cylinder perturbs  only weakly the flow when its axis is parallel to it, the perturbation is much larger in the perpendicular configuration. More precisely, the flow section may be significantly reduced
in the vicinity of the cylinder, particularly as the normalized diameter $\beta = d/h_0 \rightarrow 1$. This blockage effect forces the fluid to flow around the cylinder: it increases the  local velocity and pressure gradient and, therefore, the drag force $F$. Paragraph~\ref{lambdaperp diameter} reports measurements of $F$ for different values of $\beta$  and for various locations of the cylinder in the slit section (but always
for  $L \simeq W$). As in the parallel case (see section~\ref{force velocity}), only experiments  in which the drag force is proportional to the velocity are used (this corresponds to a Reynolds number lower than $5$).

The blockage effect discussed above is reduced when the cylinders do not span  the full width $W$ of the slit ({\it i.e.} $L < W$): then, a part of the flow  takes place between the ends of the cylinder and the side walls. This bypass effect reduces in turn the drag force $F$.
Paragraph~\ref{lambdaperp length} reports experimental measurements of the variation of $\lambda_\bot$ as a function of $L/W$
 together with the $3D$ numerical simulations which are needed to reproduce the highly three-dimensional bypass flows.

We observed that, when  perpendicular to the flow, the cylinders move
towards the middle of the cell while remaining  parallel to the plates (except for short cylinders) even for Reynolds numbers as low as $0.1$.
Such a repulsion effect has already
been  reported by~\cite{Juarez2000,Zovatto2001} and is related to inertial effects of small magnitude (in a pure viscous flow, the lift would be zero ~\cite{GHP}).
In contrast, the studies of Zovatto and Juarez~\cite{Juarez2000,Zovatto2001}  predicted an attraction by the walls due to variation of the shear in the slit gap, resulting in a shift - increasing with the Reynolds number - of the equilibrium position toward the walls. However, this phenomenon was observed only for Reynolds number larger than $80$ while the present study deals with Reynolds number  ($Re < 5$);  this effect may therefore be expected to be negligible in the present work and the  cylinders should reach (as indeed observed) an equilibrium  in the middle plane of the cell.
%%%%%%%%%%%%%%%%%%%%%%%%%%%%%%%%
  \subsection{Variation of $\lambda_\bot$ with the diameter and the location of the cylinder in the aperture}
  \label{lambdaperp diameter}
%%%%%%%%%%%%%%%%%%%%%%%%%%%%%%%%%%%
\begin{figure}[htbp]
\includegraphics[width=\W]{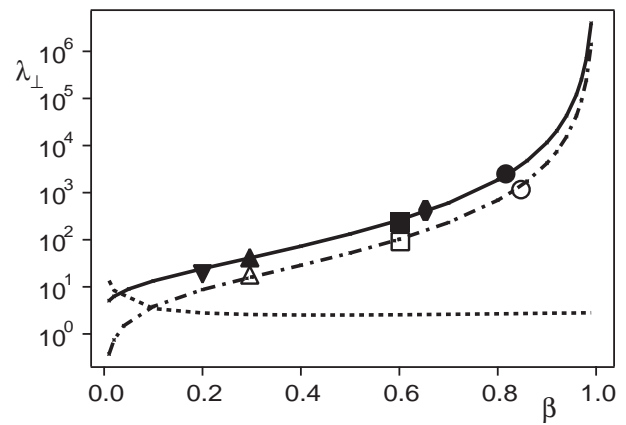}
\caption{\label{fig:fig8}Variation of $\lambda_\bot$ as a function of $\beta$ for $L \approx W$. Cylinder in contact with a wall: open symbols (experiments); dashed dotted line ($2D$ numerical computations). Cylinder in the middle plane of the cell: filled symbols (experiments);  solid line ($2D$ numerical computations, see Tab.~\ref{tab:table3}). Dotted line: ratio of the values corresponding to the solid and dashed dotted lines.}
\end{figure}
%%%%%%%%%%%%%%%%%%%%%%%%%%%%%%%%%%%%%%
In these experiments, the length of the cylinder is chosen as close as possible to the width of the cell
($0.93<L/W<1$)  in order to minimize the lateral bypass flow (this point is discussed in detail in the next subsection).
In this case, Figure \ref{fig:fig8} displays the variation of $\lambda_\bot$ as a function of $\beta$,
both for cylinders located  in the middle plane of the cell and in contact with the front wall; this latter case is
achieved experimentally by inserting a magnetic wire in the cylinder and attracting it with a magnet.

Here, too, the experimental results and the $2D$ numerical simulations agree (to better than $10\%$). Experimental values of $\lambda_\bot$ up to $2500$ are measured while, as shown in Fig. \ref{fig:fig6}, the maximal value of $\lambda_{||}$ is about $18$; such large values and the divergence of $\lambda_\bot$ when $\beta \rightarrow 1$ are due to the small gap left for the flow between the cylinder and the front walls. As this gap decreases, the pressure drop corresponding to a given constant  flow rate  rises strongly, leading to a sharp increase of the drag force $F$ on the rod.

When the axis of the cylinder is in the middle plane of the cell, the minimum hydraulic aperture of each of the two spaces between the cylinder
and the nearest wall is $h_0(1-\beta)/2$ and the flows in both flow paths add up. If the rod is displaced from its equilibrium position
and touches one of the walls along its full length, there is only one flow path of minimum hydraulic aperture $h_0(1 - \beta)$
({\it i.e.} twice the previous one). In the lubrication limit ($\beta \rightarrow 1$), the pressure drop $\Delta p$ for a given flow rate $q$
varies as the power $-5/2$ of the hydraulic aperture as shown by Eq.~(\ref{analytical value of lambdaperp}). Even taking into account the fact
that the flow takes place in a single channel, this variation with the aperture is fast enough so that the drag force is lower
in the configuration in which the cylinder touches the wall: more quantitatively, from Eq.~(\ref{analytical value of lambdaperp wall}), the coefficient $\lambda_\bot$ decreases by a factor $2\sqrt{2}$ from its value in the centered position. This difference between the drag forces measured in these two locations of the cylinder is indeed observed experimentally as shown in Fig.~\ref{fig:fig8} (dashed line).

The case $L \approx W$ seems therefore to correspond well to the $2D$ configuration  studied numerically~\cite{Dvinsky1987,benRichou2004,Champmartin2007}; however, when the cylinder is shorter than the width $W$ of the slit, flow may be perturbed (particularly when $\beta \rightarrow 1$) so that the $2D$ approximations do not reproduce well the observations. These effects will now be investigated.
%%%%%%%%%%%%%%%%%%%%%%%%%%%%%%%%%%%%%
\subsection{Variation of $\lambda_\bot$ with the length of the cylinder.}
\label{lambdaperp length}
%%%%%%%%%%%%%%%%%%%%%%%%%%%%%%%%%%%%%
In this section,  the  variation of the transverse drag coefficient $\lambda_\bot$ is studied as a function of the
 normalized length $L/W $ characterizing the lateral confinement.

%%%%%%%%%%%%%%%%%%%%%%%%%%%%%%%%
\begin{figure}[htbp]
\includegraphics[width=\W]{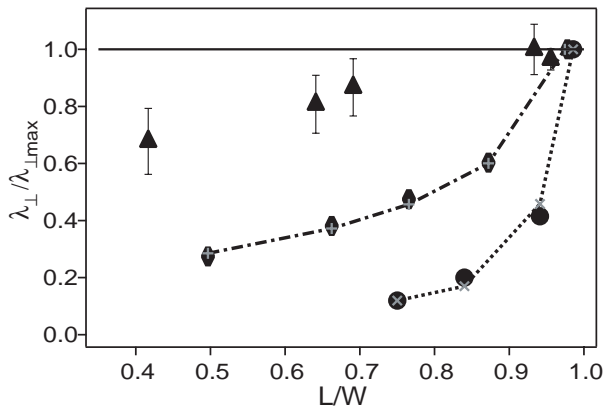}
\caption{\label{fig:fig9}Variation of the normalized drag coefficient $\lambda_{\bot}/\lambda_\bot^{max}$  as a function of the normalized length of the cylinder $L/W$, for cylinders of different diameters located in the middle plane of the cell.
($\blacktriangle$), ($\blacklozenge$) and ($\bullet$) symbols correspond respectively to
experiments for $\beta = 0.31$, $0.65$ and $0.83$ (see Tab.~\ref{tab:table2} for more details).
Dashed-dotted and dashed line: $3D$ numerical results obtained for $\beta = 0.65$ and $0.83$.}
\end{figure}
%%%%%%%%%%%%%%%%%%%%%%%%%%%%%%%%%%%%%%%
The experimental variations of $\lambda_\bot$ with $W$ for  three cylinders of  different diameters $d$ have first been compared.
For each cylinder, the maximum value $\lambda_\bot^{max}$ of $\lambda_\bot$ is reached when $L/W$ is close to $1$. In this case, the  values of $\lambda_\bot^{max}$ are close to the predictions of $2D$ simulations (see Fig.\ref{fig:fig8}); small fluctuations are observed and are likely due to experimental uncertainties (inhomogeneities of the cell aperture and cell roughness for instance).

Figure~\ref{fig:fig9} displays the experimental  variation (symbols) of the normalized ratio $\lambda_\bot /\lambda_\bot^{max}$ with $L/W $ in the range $0.42 \le L/W \le 0.99$.
As $L/W$ decreases away from $1$, the ratio $\lambda_\bot/\lambda_\bot^{max}$ becomes significantly lower than $1$; this variation occurs earlier and is particularly strong when $\beta$ is large: for instance, for $\beta = 0.83$, $\lambda_\bot/\lambda_\bot^{max}$ decreases by $60\,\%$  for a small reduction of $L/W$ by $5\,\%$.
This sharp variation is due to the partial diversion of the flow into the free space between the ends of the rod and the side walls which, in turn, reduces the drag force $F$.  As  observed in Fig.~\ref{fig:fig9}, this bypassing effect is particularly strong when the clearance at the end of the cylinder is large (low values of $L/W$) while the interval between the surface of the cylinder and the two parallel walls is small (when $\beta \rightarrow 1$).

In order to predict numerically the variations of $\lambda_\bot /\lambda_\bot^{max}$, the full $3D$ flow velocity field ${\bf v}(x,y,z)$
must be determined, and not only  the components $v_y(y,z)$ and $v_z(y,z)$ as before. This has been achieved by using the $3D$ version of the Freefem program (see Sec.\ref{Numerical method}).
As a validation test, we consider first the special situation in which the
 length of the cylinder is equal to the width of the cell  ($L = W$).
If, in addition, a perfect slip condition is used for the side walls, then
the $3D$ and $2D$ simulations are equivalent and similar results should be obtained.
Actually, $3D$ simulations predict values of $\lambda_\bot$ larger than $2D$ ones by
 $20\,\%$ or less: this difference is likely due to  computational limitations related to
the minimal practical value of the mesh size.

These $3D$ simulations were then performed for different values of  $L/W$ and for two
different normalized diameters $\beta$ corresponding to actual experiments.
The dotted and dashed-dotted lines in Fig~\ref{fig:fig9} connect the data points corresponding to
the ratios $\lambda_\bot/\lambda_\bot^{max}$  obtained from these simulations (the values
of $\lambda_\bot^{max}$ are those obtained in the validation simulations for $L=W$).
The experimental and numerical variations of $\lambda_\bot/\lambda_\bot^{max}$ with $L/W$ are
in very good agreement. This confirms that, in this geometry, the variations of the drag force with $L/W$
reflect  $3D$ modifications of the flow structure and cannot be accounted for by
$2D$ models.

\begin{figure} [htbp]
\includegraphics[width=\W]{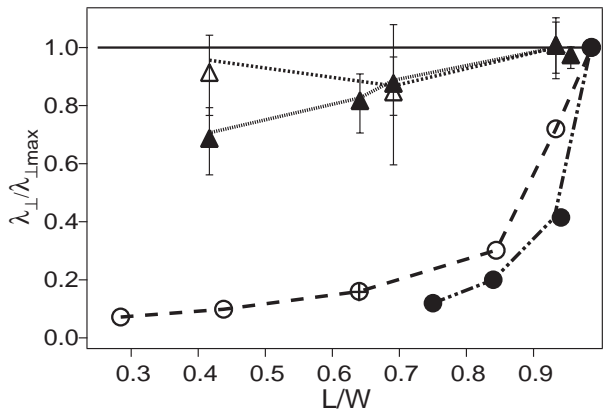}
\caption{\label{fig:fig10}Variation of the normalized drag coefficient $\lambda_{\bot}/\lambda_\bot^{max}$
as a function of the normalized length $L/W$ for two cylinders of different diameters
 located either in the middle plane of the cell (filled symbols) or near a wall (open symbols).
($\triangle$,$\blacktriangle$) and ($\circ$, $\bullet$) symbols correspond respectively to $\beta = 0.31$ and $0.83$
(see Tab.~\ref{tab:table2}). Lines are guides for eyes.}
\end{figure}
%%%%%%%%%%%%%%%%%%%%%%%%%%%%%%%%%%%%%%%
The variation of $\lambda_\bot$ with the distance of the rod from the middle plane of the cell has also been investigated:
as shown in Sec.~\ref{lambdaperp diameter}, the drag force should be lower when the rod is in contact with one of
the front walls than when it is located halfway between them.
We compared  the experimental variations of $\lambda_{\bot}/\lambda_\bot^{max}$ with $L/W$ in both configurations: the
results obtained for  two values of the normalized diameter $\beta$ are displayed as symbols  in Fig.~\ref{fig:fig10}.
The drop of the coefficients when $L$ decreases is more pronounced for a cylinder halfway between the walls.
This can be explained by the different relative magnitude of the hydraulic impedance of the  flow paths between the ends of
the rod and the side walls and of the direct paths between the rod and the front walls.

Finally, the influence of the viscosity has been investigated by comparing  the results of experiments using identical parameters but fluids with two different viscosities: $\eta=50 \,\mathrm{mPa.s}$ ($+$) and $\eta=30 \,\mathrm{mPa.s}$ ($\circ$). The points coincide, which confirms that the drag force is proportional to the viscosity.
%%%%%%%%%%%%%%%%%%%%%%%%%%%%%%%%%%
\section{Conclusion}
\label{Conclusion}
%%%%%%%%%%%%%%%%%%%%%%%%%%%%%%%%%%%%%%%%%%%%
The experiments and numerical simulations reported in the present paper have allowed one to determine the
influence of confinement effects on the drag force $F$ on a static rigid cylinder in a viscous flow inside
a rectangular slit. Significantly different results have been obtained in the cases of cylinders with their length
parallel and perpendicular to the mean flow, although, in both cases, $F$ is proportional to the mean velocity and
the viscosity in the range investigated and can, therefore, be characterized by  drag coefficients $\lambda_{||}$
and $\lambda_\bot$.

In the parallel case, $\lambda_{||}$ increases linearly with the confinement parameter $\beta=d/h_0$ but does not
diverge for $\beta= 1$; $\lambda_\bot$ increases much faster with $\beta$ and diverges near $\beta =1$ (due to the
 blockage of the flow) when the length $L$ of the cylinder is close to the width $W$ of the slit.
$2D$ numerical simulations in planes respectively perpendicular (for $\lambda_{||}$) and parallel (for $\lambda_\bot$ and
$L \simeq W$) to the mean flow   reproduce well the results obtained in these two cases. In the perpendicular case
analytical model based on the lubrication also provides a good agreement, still for $L \simeq W$ and for a strong enough
confinement  ({\it{i.e.}} for $\beta>0.2$).

When the cylinders are shorter than the width $W$ of the slit, a bypass flow  appears in the space between the
 between the edges of the cylinder and the side walls of the slit: this reduces the direct flow in the gap between
 the front walls and the cylinders.  This effect is particularly strong when the  confinement parameter $\beta$ is close
 to $1$; it results  in a sharp decrease of the coefficient $\lambda_\bot$.  $3D$ numerical simulations are needed in order
 to predict quantitatively this effect.

The present experiments also provided evidence for an inertial lift force in the case of cylinders perpendicular to the flow direction.
This force was observed for Reynolds numbers as low as $0.1$ and kept the cylinders in the middle plane of the model: this may explain recent
 observation of the depinning of fibers trapped inside fractures \cite{Dangelo2008}. This observation may be contrasted with numerical
 simulations~\cite{Juarez2000,Zovatto2001} which predict an opposite effect, {\it{i.e.}} a force pushing the cylinder towards the walls. This
 latter force should however  appear only at larger Reynolds numbers: further studies are needed to investigate these issues.

The present study dealt only with motionless rigid cylinders inside a viscous flow. Extending these studies to the measurement of forces on
moving cylindrical objects, both rigid and flexible, will make the results applicable to the motion of freely swimming microorganisms \cite{Diluzio2005}.
%%%%%%%%%%%%%%%%%%%%%%%%%%%%%%
\begin{acknowledgments}
We thank R. Pidoux for realizing the
experimental setup and A. Aubertin for the data acquisition program. We also thank
S. Del Pino for providing us with help in the use of FreeFEM3D and J. Hinch for helpful suggestions and discussions.
\end{acknowledgments}
%\begin{thebibliography}{99}
%\bibliographystyle{jasanum.bst}
%\bibliography{fibre}
%\end{thebibliography}

\end{document}